\documentclass[12pt]{article}
\usepackage{graphicx}
\usepackage{amsmath,amssymb}
\usepackage{colortbl}
\usepackage{multirow} 
\usepackage[abs]{overpic} 

\newcommand\pubnumber{}
\newcommand\pubdate{\today}

\def\utokyo{The University of Tokyo \\ 
Department of Physics, Graduate School of Science \\ 
7-3-1 Hongo, Bunkyo-ku Tokyo 113-0033 Japan}

\def\Title#1{\begin{center} {\Large #1 } \end{center}}
\def\Author#1{\begin{center}{ \sc #1} \end{center}}
\def\Address#1{\begin{center}{ \it #1} \end{center}}

\newcommand\pubblock{\rightline{\begin{tabular}{l} \pubnumber\\
         \pubdate  \end{tabular}}}
\newenvironment{Abstract}{\begin{quotation}  }{\end{quotation}}
\newenvironment{Presented}{\begin{quotation} \begin{center} 
             PRESENTED AT\end{center}\bigskip 
      \begin{center}\begin{large}}{\end{large}\end{center} \end{quotation}}

\begin{document}
\begin{titlepage}
\pubblock

\vfill
\Title{Tau lifetime and decays}
\vfill
\Author{Denis Epifanov}
\Address{\utokyo}
\vfill
\begin{Abstract}
Recent results of a high-statistics study of
$\tau$ lepton properties and decays at $B$ factories are reviewed. 
We discuss measurements of $\tau$ lifetime, branching fractions, 
and spectral functions for several hadronic $\tau$ decay modes with $K^0_S$. 
Results of a search for lepton flavor violating $\tau$ decays 
as well as CP symmetry violation are briefly discussed.
\end{Abstract}
\vfill
\begin{Presented}
Flavor Physics and CP Violation (FPCP-2014), \\
Marseille, France, May 26-30, 2014
\end{Presented}
\vfill
\end{titlepage}
\def\thefootnote{\fnsymbol{footnote}}
\setcounter{footnote}{0}

\section{Introduction}

The world largest statistics of $\tau$ leptons collected 
at $e^+ e^-$ $B$ factories (Belle \cite{belle,kekb} and 
$BABAR$ \cite{Aubert:2001tu,Kozanecki:2000cm}) and 
LHCb \cite{Alves:2008zz} opens new era in the precision tests of the Standard Model (SM). 
Basic tau properties like lifetime, mass, couplings, 
electric dipole moment (EDM), anomalous magnetic dipole moment 
and other are introduced as free parameters in the theory, 
or they can be calculated in the SM. Hence the former 
parameters should be measured experimentally as precise as 
possible, while the latter ones provide the unique 
possibility to test SM and search for the effects 
of New Physics (NP). An essential progress has been made 
in the study of the main $\tau$ properties at Belle and $BABAR$, 
namely lifetime \cite{Belous:2013dba,Lusiani:2005sy}, 
mass \cite{Abe:2006vf,Aubert:2009ra}, EDM \cite{Inami:2002ah}, 
coupling constant ratios \cite{Aubert:2009qj} have been 
measured with the best or competitive to the world best 
accuracies \cite{Beringer:1900zz}. 

In the SM $\tau$ decays due to the charged weak interaction 
described by the exchange of $W^{\pm}$ with a pure vector 
coupling to only left-handed chirality fermions. There are 
two main classes of tau decays: leptonic decays\footnote{Unless specified
otherwise, charge conjugate decays are implied throughout the paper.} 
($\tau^-\to \ell^-\bar{\nu_{\ell}}\nu_{\tau}$, 
$\tau^-\to \ell^-\bar{\nu_{\ell}}\nu_{\tau}\gamma$, 
$\tau^-\to \ell^- \ell'^+\ell'^- \bar{\nu_{\ell}}\nu_{\tau}$;~
$\ell,\ell' = e,\mu$), and hadronic decays. 
Leptonic decays provide very clean laboratory to probe 
electroweak couplings \cite{Fetscher:1993ki}, which is 
complementary or competitive to the precision studies of muon 
in the experiments with muon beams \cite{Kuno:1999jp}. Plenty 
of NP models can be tested and constrained in the precision 
studies of the dynamics of $\tau$ decays with leptons \cite{Haber:1978jt}-\cite{McKeen:2011aa}. 

Hadronic decays of $\tau$ offer unique tools for the precision 
study of low energy QCD \cite{Pich:1997ym}. The hadronic system 
is produced from the QCD vacuum via decay of the $W^-$ boson 
into $\bar{u}$ and $d$ quarks (Cabibbo-allowed decays) 
or $\bar{u}$ and $s$ quarks (Cabibbo-suppressed decays). 
As a result the decay amplitude can be factorized into a 
purely leptonic part including the $\tau$ and $\nu_{\tau}$ 
and a hadronic spectral function. 
Various decay modes are interesting to study precisely the structure of the 
hadronic spectral functions~\cite{Kuhn:1992nz} and measure precisely parameters 
of the intermediate states, testing the Wess-Zumino-Witten anomaly~\cite{Wess:1971yu}, 
chiral theory~\cite{Pich:1987qq,Pich:1995bw}, and relations to $e^+e^-$ cross sections 
following from the conservation of the vector current~\cite{Eidelman:1990pb}. 
Measurement of the inclusive hadronic spectral function of Cabibbo-allowed 
decays is important for the precision determination of $\alpha_s$ \cite{Braaten:1991qm}, 
while the inclusive strange hadronic spectral function is used to evaluate 
$s$-quark mass and $V_{us}$ element of Cabibbo-Kobayashi-Maskawa (CKM) 
quark flavor-mixing matrix \cite{Gamiz:2004ar}. Recently lots of important results in the hadronic 
sector of $\tau$ physics have been obtained at Belle and $BABAR$ \cite{Bevan:2014iga}. 

In the leptonic sector CP symmetry violation (CPV) is strongly suppressed 
in the SM ($A_{\rm SM}^{\rm CP}\lesssim 10^{-12}$) leaving enough room to 
search for the effects of NP~\cite{Delepine:2005tw}. Of particular interest are 
strangeness changing Cabibbo-suppressed hadronic $\tau$ decays, in which 
large CPV could appear from a charged scalar boson exchange~\cite{Weinberg:1976hu}-\cite{Kiers:2008mv}. 
 
Probabililty of lepton flavor violating (LFV) decays of charged leptons is 
extremely small in the Standard Model 
(for example ${\cal B}(\tau\to\ell\nu) \sim \Delta m^4_\nu / m^4_W < 10^{-54}$ \cite{Deppisch:2012vj}). 
Many models beyond the SM predict LFV decays with the branching fractions 
up to $\lesssim 10^{-7}$ \cite{Ellis:2002fe}-\cite{CorderoCid:2005gr}. 
As a result observation of LFV is a clear signature of New Physics. 
$\tau$ lepton is an excellent laboratory to search for the LFV 
decays due to the enhanced couplings to the new particles as well as large 
number of LFV decay modes. Study of different $\tau$ LFV decay modes allows 
one to test various NP models. Huge statistics collected by Belle and $BABAR$ 
was used in the searches for 48 LFV $\tau$ decays, the upper limits on the 
branching fractions for the most of LFV modes approach $10^{-8}$ level, which 
allows one to constrain the parameter spaces of many NP models \cite{Bevan:2014iga}. 
Recently LHCb collaboration performed results of the search for LFV $\tau^-\to\mu^-\mu^+\mu^-$, 
lepton number (LNV) and barion number violating (BNV) $\tau$ decays with proton 
$\tau^-\to p\mu^-\mu^-,~\bar{p}\mu^+\mu^-$ at the Large Hadron Collider \cite{Aaij:2013fia}. 

\section{Measurement of $\tau$ lifetime at Belle}
\subsection{Tests of lepton universality} 
Lepton universality in the charged lepton sector of the SM 
is the fundamental assumption about lepton flavor-independent 
structure of the charged weak interaction. It is introduced in 
the theory as an equality of the coupling constants for $e^-$, 
$\mu^-$ and $\tau^-$: $g_e=g_\mu=g_\tau$. 
This universality can be experimentally tested by comparing 
the rates of the leptonic decays: $\tau^-\to e^-\bar{\nu}_e\nu_\tau$, 
$\tau^-\to \mu^-\bar{\nu}_\mu\nu_\tau$ and $\mu^-\to e^-\bar{\nu}_e\nu_\mu$. 
The total decay width with electroweak radiative corrections of 
lepton $L^-$ ($L=\mu,~\tau$) reads \cite{Marciano:1988vm}: 
\begin{equation}
\Gamma(L^-\to\ell^-\bar{\nu}_{\ell}\nu_L(\gamma))=\frac{{\cal B}(L^-\to\ell^-\bar{\nu}_{\ell}\nu_L(\gamma))}{\tau_L}=\frac{g^2_L g^2_\ell}{32M^4_W}\frac{m^5_L}{192\pi^3}F_{\rm corr}(m_L,m_\ell), 
\end{equation}
\begin{equation}
F_{\rm corr}(m_L,m_\ell)=f(x)\biggl(1+\frac{3}{5}\frac{m^2_L}{M^2_W}\biggr)\biggl(1+\frac{\alpha(m_L)}{2\pi}\biggl(\frac{25}{4}-\pi^2\biggr)\biggr), 
\end{equation}
\begin{equation}
f(x)=1-8x+8x^3-x^4-12x^2\ln{x},~x=m_\ell/m_L, 
\end{equation}
where $m_L$($m_\ell$) and $g_L$($g_\ell$) are mass and coupling constant of initial (final) lepton, 
$\tau_L$ is lifetime of initial lepton, $M_W$ and $\alpha(m_L)$ are $W^-$ boson mass and fine-structure 
constant at the energy scale of $m_L$. 
Taking into account that \cite{Beringer:1900zz}:    
\[ {\cal B}(\mu^-\to e^- \bar{\nu}_e\nu_\mu(\gamma))={\cal B}(\mu^-\to e^- \bar{\nu}_e\nu_\mu)+\]
\begin{equation}
{\cal B}(\mu^-\to e^- \bar{\nu}_e\nu_\mu\gamma)+{\cal B}(\mu^-\to e^- \bar{\nu}_e\nu_\mu e^+ e^-)=1, 
\end{equation} 
the $g_\tau /g_e$ and $g_\tau /g_\mu$ ratios of the coupling constants can be extracted: 
\begin{equation}
\frac{g_\tau}{g_e}=\sqrt{{\cal B}(\tau^-\to\mu^-\bar{\nu}_\mu\nu_\tau(\gamma))\frac{\tau_\mu}{\tau_\tau}\frac{m^5_\mu}{m^5_\tau}\frac{F_{\rm corr}(m_\mu,m_e)}{F_{\rm corr}(m_\tau,m_\mu)}}, 
\label{gtau:ge} 
\end{equation} 
\begin{equation} 
\frac{g_\tau}{g_\mu}=\sqrt{{\cal B}(\tau^-\to e^-\bar{\nu}_\mu\nu_\tau(\gamma))\frac{\tau_\mu}{\tau_\tau}\frac{m^5_\mu}{m^5_\tau}\frac{F_{\rm corr}(m_\mu,m_e)}{F_{\rm corr}(m_\tau,m_e)}}. 
\label{gtau:gmu} 
\end{equation} 
As it is seen from Eq.~\ref{gtau:ge} and \ref{gtau:gmu} precise measurement of the branching fraction 
of leptonic $\tau$ decay, $\tau$ mass, and $\tau$ lifetime are necessary for the accurate tests of 
lepton universality. 
According to the last report from Heavy Flavor Averaging Group (HFAG) \cite{Amhis} lepton 
universality was confirmed in the $g_\tau /g_e$ and $g_\tau /g_\mu$ ratios with the accuracy 
of about 0.2\%: 
\[ g_\tau/g_e=1.0024\pm 0.0021,~g_\tau/g_\mu=1.0006\pm 0.0021 \] 
However, recently LEP Electroweak Collaboration published results on the test of 
lepton universality in W-boson decays \cite{Schael:2013ita}, and the ratio of the 
branching fraction of $W^-$-boson decay to $\tau^-\bar{\nu}_\tau$ to the average 
branching fraction of $W^-$-boson decay to $\mu^-\bar{\nu}_\mu$ and $e^-\bar{\nu}_e$
was found to be 2.6 standard deviations away from unity:
\[\frac{2{\cal B}(W^-\to\tau^-\bar{\nu}_\tau)}{{\cal B}(W^-\to\mu^-\bar{\nu}_\mu) + {\cal B}(W^-\to e^-\bar{\nu}_e)}=1.066\pm 0.025.\]
So, the improvement of the accuracy of the test of lepton universality with 
leptonic decays of $\tau$ is still rather actual task. The reduction of the 
uncertainty of tau lifetime to the negligible level will allow one to test 
lepton universality in $g_\tau /g_e$ and $g_\tau /g_\mu$ ratios with the 
accuracy of about 0.1\% even with the current values of the $\tau$ mass 
and branching fraction uncertainties. 
 
\subsection{Measurement of $\tau$ lifetime at Belle} 

This analysis \cite{Belous:2013dba} is based on the statistics with the luminosity integral of 
$\int Ldt=711$~fb$^{-1}$ collected with the Belle detector  
at the KEKB asymmetric-energy $e^+ e^-$ collider operating 
at the $\Upsilon(4S)$ resonance and 60~MeV below. The data sample comprises 
653$\times 10^6~\tau^+\tau^-$ pairs. Events where both taus decay to three 
charged pions and neutrino were selected: $e^+e^-\to\tau^+\tau^-\to (\pi^+\pi^+\pi^-\bar{\nu}_{\tau},~\pi^+\pi^-\pi^-\nu_{\tau})$ 
(or shortly $(3\pi)^+ - (3\pi)^-$). 
\begin{figure}[hbtp]
\centering 
\includegraphics[width=0.4\textwidth]{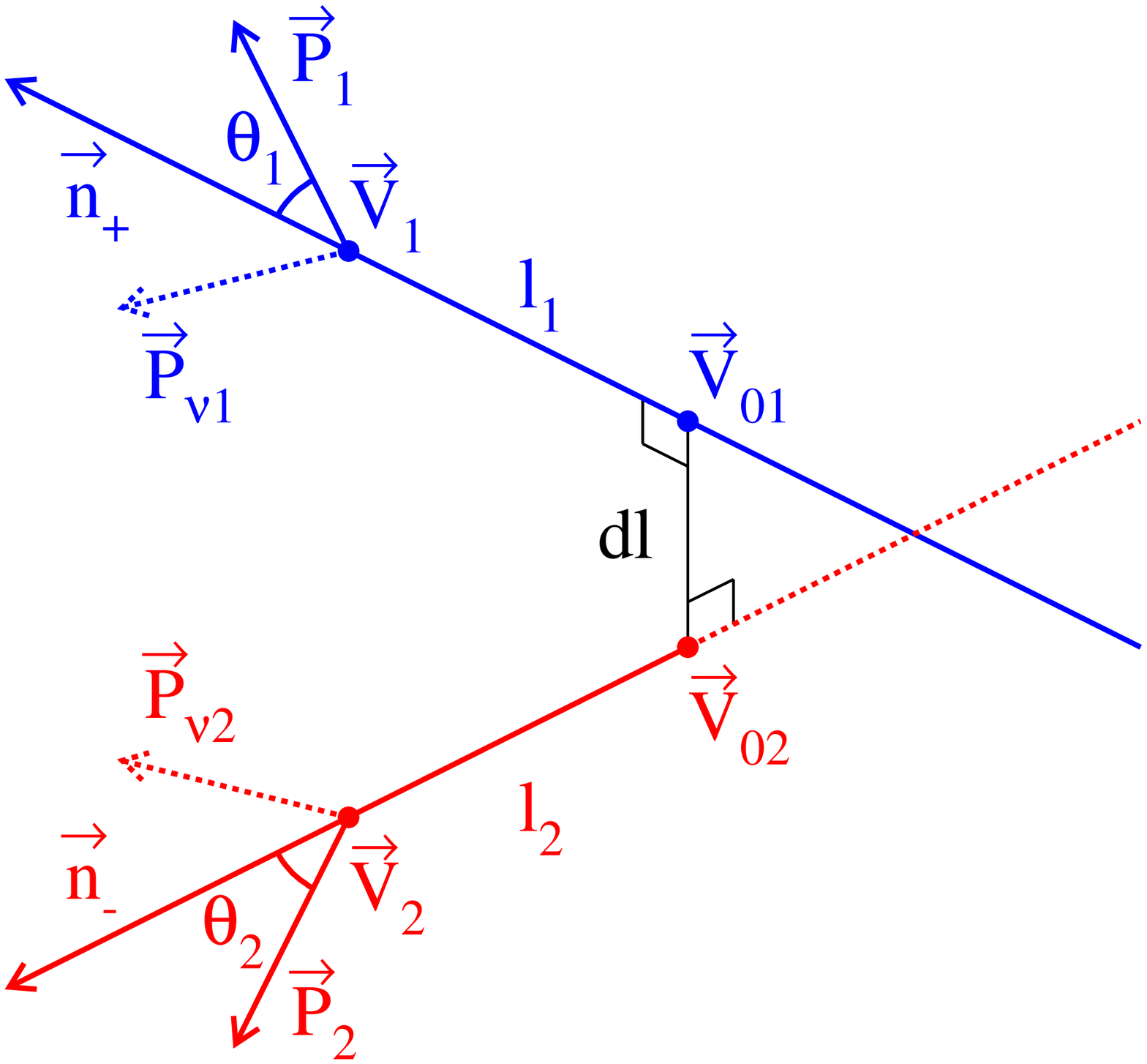} 
\hspace{5mm}
\includegraphics[width=0.4\textwidth]{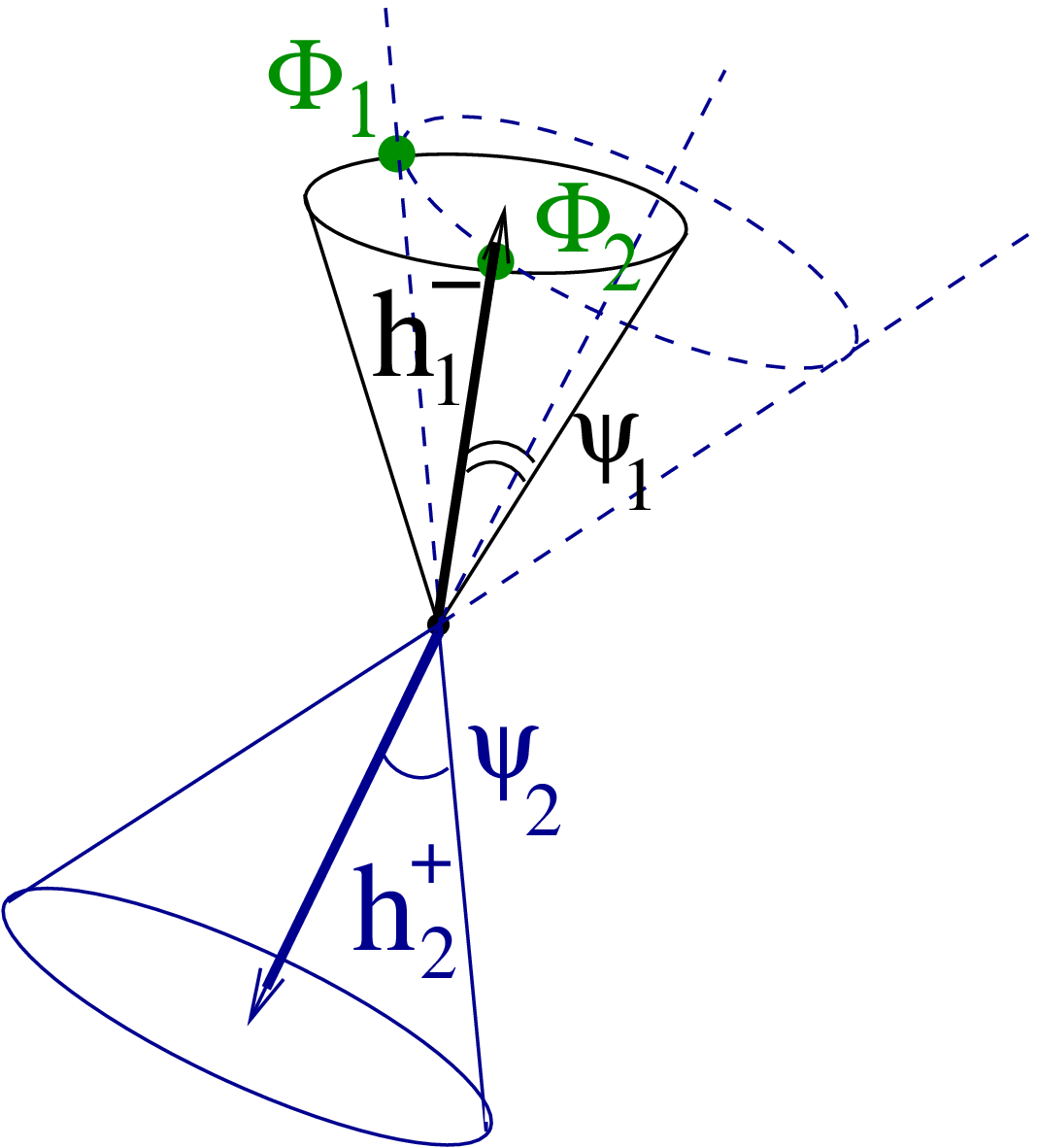}
\\
\parbox[t]{0.48\textwidth}{\caption{The scheme of $(3\pi)^+ - (3\pi)^-$ event in laboratory frame.}\label{ctaumeth}}
\parbox[t]{0.48\textwidth}{\caption{The scheme of $(3\pi)^+ - (3\pi)^-$ event 
in c.m.s.~$\cos{\psi_{1,2}} = \frac{2E_{\tau}E_{h_{1,2}}-M^2_{\tau}-m^2_{h_{1,2}}}{2p_{\tau}p_{h_{1,2}}}$, 
$h^-_1 = (3\pi)^-$, $h^+_2 = (3\pi)^+$.}\label{taudir}} 
\end{figure}
At the asymmetric-energy $e^+ e^-$ collider the angle between 
$\tau^+$ and $\tau^-$ in laboratory frame is smaller than $180^\circ$, 
so the $\tau^+\tau^-$ production point can be calculated from the 
intersection of two trajectories defined by the $\tau$-lepton 
decay vertices and their momentum directions, see Fig.~\ref{ctaumeth}.
The position of beam interaction point is not needed at all in this method. 
$\tau$ momentum direction is determined with the two-fold ambiguity in the 
center-of-mass system (c.m.s.), for the analysis average axis is used, 
see Fig.~\ref{taudir}. 
In this method lifetimes of $\tau^-$ and $\tau^+$ can be measured 
separately to test CPT symmetry conservation. 
  
The following criteria were applied to select signal $(3\pi)^+ - (3\pi)^-$ 
events: six charged pions with zero net charge and no 
other tracks are found; the thrust value in the c.m.s.~is 
greater than 0.9; three pions (triplet) with $\pm 1$ net charge 
in each hemisphere, separated by the plane perpendicular 
to the thrust axis in the c.m.s.; there are no additional 
$K^0_S$, $\Lambda$, and $\pi^0$ candidates; absolute value 
of the transverse momentum of the $6\pi$ system is greater 
than 0.5~GeV/$c$; the mass of $6\pi$ system should 
satisfy the requirement $4$~GeV/$c^2<M(6\pi)<10.25$~GeV/$c^2$; the 
pseudomass of each triplet of pions is $\sqrt{M^2_h+2(E_{\rm beam}-E_h)(E_h-P_h)}<1.8$~GeV/$c^2$, 
$h=(3\pi)^-,~(3\pi)^+$; each triplet vertex-fit quality fulfill $\chi^2<20$; 
the distance ($dl$) of the closest approach of $\tau^-$ and $\tau^+$ 
trajectories in laboratory frame satisfy $dl<0.03$~cm. 
Finally 1148360 events were selected with the background contamination 
of about 2\%, the dominant background comes from the continuum 
$e^+ e^-\to q\bar{q}$ ($q=$u,~d,~s) events.  
  
\begin{figure}[hbtp]
\centering 
\includegraphics[width=0.4\textwidth]{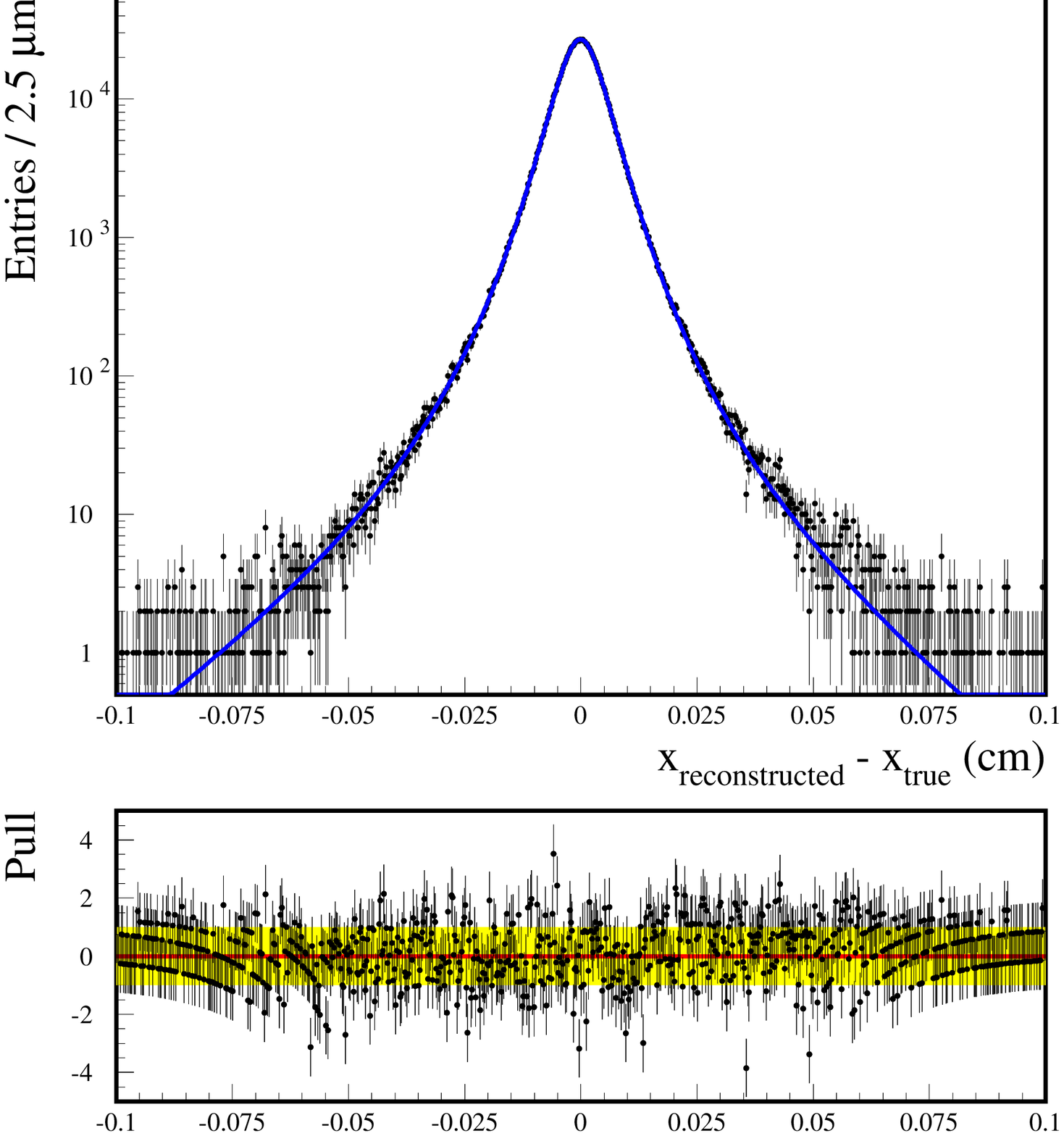} 
\hspace{5mm}
\includegraphics[width=0.4\textwidth]{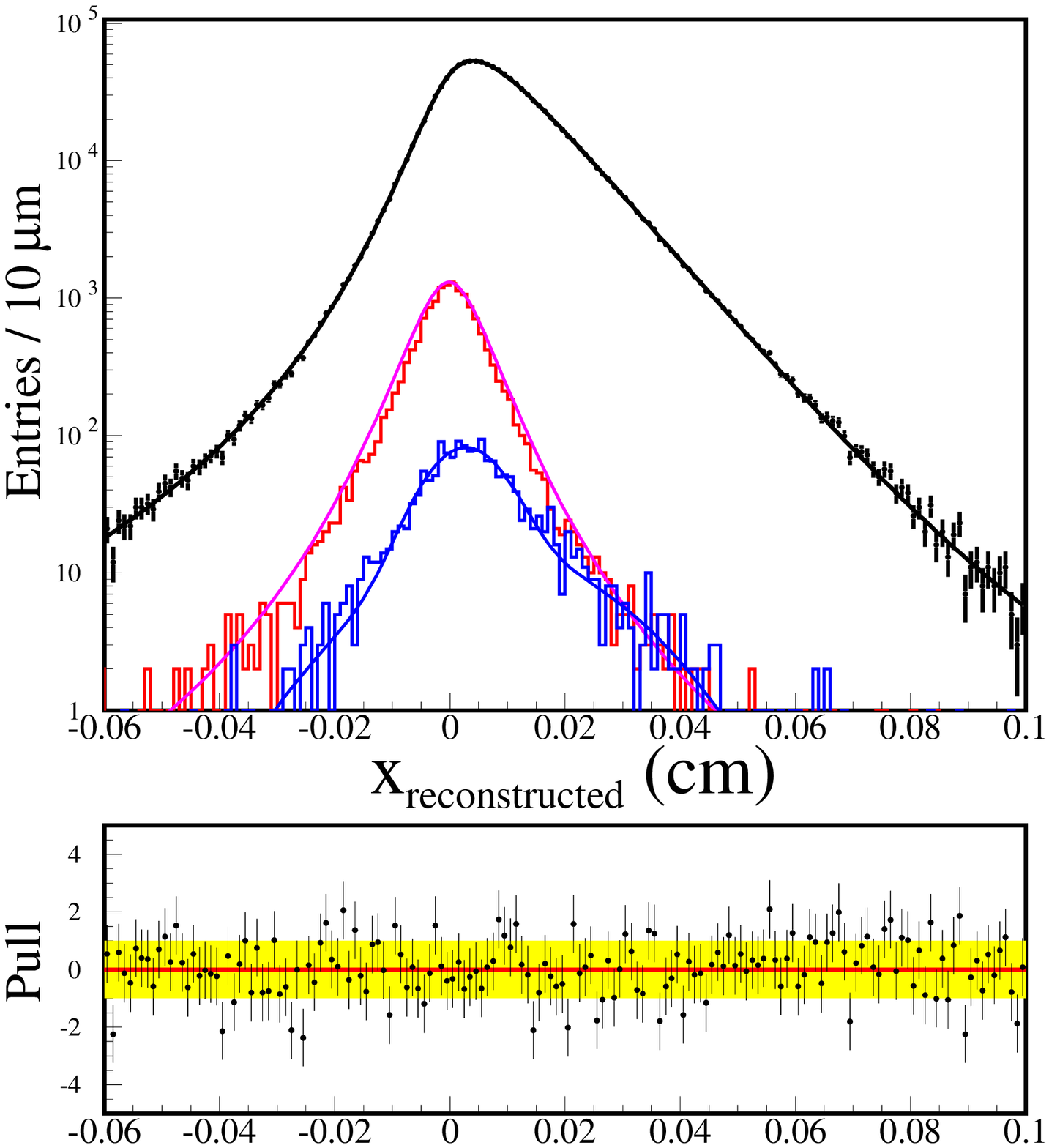}
\\
\parbox[t]{0.48\textwidth}{\caption{Distribution of the difference between the reconstructed 
and true $\tau$ decay length values, line shows result of the fit to Eq.~\ref{resofun}. 
Distribution of the residuals for the fit is also shown.}\label{ctaures}} 
\parbox[t]{0.48\textwidth}{\caption{Distribution of the experimentally measured decay length 
(points with errors), result of the fit is shown by black solid line. Red histogram and solid 
line show the MC prediction and parametrisation for the $e^+ e^-\to q\bar{q}$ ($q=$u,~d,~s) 
and two-photon backgrounds, blue histogram and solid line show the MC prediction and 
parametrisation for the $e^+ e^-\to q\bar{q}$ ($q=$c,~b) background. Distribution of the 
residuals for the fit is also shown.}\label{ctaufit}} 
\end{figure}

The probability density function (p.d.f.) for the measured 
$\tau$ decay length distribution is written in the form: 
\begin{equation} 
{\cal P}(x)={\cal N}\int e^{-x'/\lambda_{\tau}}R(x-x';\vec{P})dx'+{\cal N}_{\rm uds}R(x;\vec{P})+B_{\rm cb}(x), 
\end{equation} 
where $x = \ell/(\beta_{\tau}\gamma_{\tau})$ is normalized $\tau$ decay length, ${\cal N}$ is normalisation 
constant, $\lambda_{\tau}$ is estimator of $c\tau_{\tau}$ and $c\tau_{\tau}=\lambda_{\tau}+\Delta_{\rm corr}$ 
($\Delta_{\rm corr}$ is determined from MC), ${\cal N}_{\rm uds}$ is contribution of background 
from $e^+e^-\to q\bar{q}~(q=u,~d,~s)$ (predicted by MC), $B_{\rm cb}(x)$ is p.d.f. contribution to describe background from 
$e^+e^-\to q\bar{q}~(q=c,~b)$ (fixed from MC), $R(x;\vec{P})$ is detector resolution function (see Fig.~\ref{ctaures}), 
parametrized by: 
\[ R(x;\vec{P}) = (1-2.5x)\cdot\exp\biggl(-\frac{(x-P_1)^2}{2\sigma^2}\biggr),\] 
\begin{equation}  
\sigma=P_2+P_3|x-P_1|^{1/2}+P_4|x-P_1|+P_5|x-P_1|^{3/2}. 
\label{resofun}
\end{equation}
$\lambda_{\tau}$, ${\cal N}$ and $\vec{P}=(P_1,...,P_5)$ are free parameters of the fit.
From the fit of experimental data $\tau$ lifetime estimator is obtained to be 
$\lambda_{\tau}=86.53\pm 0.16$~$\mu$m, applying MC correction $\Delta_{\rm corr}=0.46$~$\mu$m 
the $\tau$ lifetime (multiplied by speed of light) value is $c\tau_{\tau}=86.99\pm 0.16$~$\mu$m., 
where the error is statistical only. The result of the fit is demonstrated in Fig.~\ref{ctaufit}.
The main sources of systematic uncertainties are summarized in Table~\ref{tab:syst}. 
\begin{table}[htbp]
\begin{center}
\begin{tabular}{l@{~}c} 
\hline
Source & $\Delta c\tau_\tau$  ($\mu$m) \\ 
\hline
Silicon vertex              & \multirow{2}*{0.090} \\
detector alignment          &       \\
Asymmetry fixing            & 0.030 \\
Fit range                   & 0.020 \\
Beam energy, ISR, FSR       & 0.024 \\
Background contribution     & 0.010 \\ 
$\tau$-lepton mass          & 0.009 \\ 
\hline
Total                       & 0.101 \\ 
\hline
\end{tabular}
\caption{Systematic uncertainties of $c\tau_\tau$.}\label{tab:syst} 
\end{center} 
\end{table} 
\begin{figure}[hbtp]
\centering 
\includegraphics[width=0.5\textwidth]{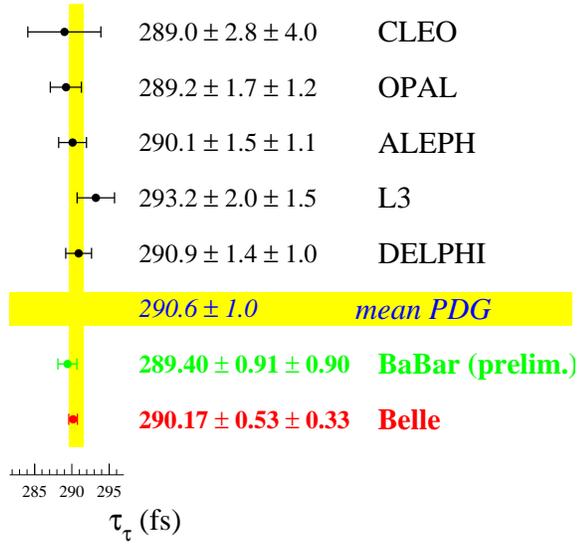} 
\caption{Summary of $\tau$ lifetime measurements.}\label{ctauresul} 
\end{figure}
The obtained results for the product of the lifetime and speed of light and for the lifetime are: 
\[ c\tau_{\tau} = (86.99 \pm 0.16({\rm stat}) \pm 0.10({\rm syst}))~\mu{\rm m.} \] 
\[ \tau_{\tau} = (290.17 \pm 0.53({\rm stat}) \pm 0.33({\rm syst}))~{\rm fs.}   \] 
Belle result on $\tau_{\tau}$ and previous measurements are shown in Fig.~\ref{ctauresul}. 
For the first time the upper limit on the relative $\tau_\tau$ 
difference between $\tau^+$ and $\tau^-$ was measured to be: 
\[ |\tau_{\tau^+}-\tau_{\tau^-}|/\tau_\mathrm{average}<7.0\times10^{-3}\mathrm{\, \, at\, \, 90\%\,\, CL.} \] 
With the new $\tau_\tau$ Belle result the $g_\tau /g_e$ and $g_\tau /g_\mu$ 
ratios were recalculated to revise lepton universality: 
\[ g_\tau/g_e=1.0031\pm 0.0016,~g_\tau/g_\mu=1.0013\pm 0.0016. \] 
It is seen that the uncertainty of the ratios was improved by a factor 
of about 1.3 in comparison with the last HFAG result \cite{Amhis}, 
and now the $g_\tau / g_e$ ratio is almost 2 standard deviations away from unity. 

\section{Hadronic $\tau$ decays with $K^0_S$ at Belle} 
\subsection{Measurement of the branching fractions} 
The analysis \cite{Ryu:2014vpc} is based on the data sample with the 
luminosity integral of ${\cal L}=669$~fb$^{-1}$ which comprises 
615 million $\tau^+\tau^-$ events. 
One inclusive decay mode $\tau^-\to K^0_S X^-\nu_{\tau}$ 
and 6 exclusive hadronic tau decay modes with $K^0_S$ 
($\tau^-\to\pi^-K^0_S\nu_\tau$, $\tau^-\to K^-K^0_S\nu_\tau$, $\tau^-\to\pi^-K^0_SK^0_S\nu_\tau$, 
 $\tau^-\to\pi^-K^0_S\pi^0\nu_\tau$, $\tau^-\to K^-K^0_S\pi^0\nu_\tau$, $\tau^-\to\pi^-K^0_SK^0_S\pi^0\nu_\tau$) 
were studied.~After the standard Belle $\tau\tau$ preselection criteria \cite{Ryu:2014vpc} 
the following selection criteria are applied: the thrust value in the c.m.s. 
is greater than 0.9; event is separated into two hemispheres separated by the 
plane perpendicular to the thrust axis; in the tag hemisphere one-prong tau decay 
is required (decays with $e$, $\mu$ or $\pi/K(n\geq 0)\pi^0$); 
in the signal hemisphere particular hadronic final state is required; 
$K^0_S$ candidate is reconstructed from a pair of oppositely charged 
pions with $\pi^+\pi^-$ invariant mass within the range $0.485$~GeV/$c^2<M_{\pi\pi}<0.511$~GeV/$c^2$, 
decay length in the $r-\varphi$ plane satisfying $2$~cm$\leq L_{\rm{\perp}}\leq 20$~cm, and 
$z$-distance between the two helices at the $\pi^+\pi^-$ vertex position before the fit 
is required to be $\Delta Z_{1,2}<2.5$~cm; $\pi^0$ candidate is reconstructed from 
a pair of $\gamma$'s with normalized $\gamma\gamma$ invariant mass satisfying 
$-6<(m_{\gamma\gamma}-m_{\pi^{0}})/\sigma_{\gamma\gamma}<5$; charged kaon/pion 
identification parameter $\mathcal{P}_{K/\pi}=\mathcal{L}_{K}/(\mathcal{L}_{\pi}+\mathcal{L}_K)$ 
is required to be $\mathcal{P}_{K/\pi}>0.7(<0.7)$ for kaon (pion); the total energy of extra 
photons (not from the signal $\pi^0$ candidates) in laboratory frame should satisfy 
$\sum E_{\rm{\gamma}}^{\rm{LAB}}< 0.2$~GeV. 
For the inclusive mode $\tau^-\to K^0_S X^-\nu_{\tau}$ the number of 
signal events is obtained from the fit of the $\pi^+\pi^-$ invariant 
mass distribution of $K^0_S$ candidate. 
\begin{table}[htbp]
\begin{center}
\small
\begin{tabular}{@{}c@{~}|@{~}c@{~}|@{~}c@{~}c@{~}c@{~~}c@{~~}c@{~~}c@{}}
\hline 
Mode & $K^0_SX^-$ & $\pi^-K^0_S$ & $K^-K^0_S$ & $\pi^-K^0_S\pi^0$ & $K^-K^0_S\pi^0$ & $\pi^-K^0_SK^0_S$ & $\pi^-K^0_SK^0_S\pi^0$ \\ 
\hline 
$N^{\rm data}$ & {\scriptsize $397806\pm 631$} & {\scriptsize $157836\pm 541$} & {\scriptsize $32701\pm 295$} & {\scriptsize $26605\pm 208$} & {\scriptsize $8267\pm 109$} & {\scriptsize $6684\pm 96$} & {\scriptsize $303\pm 33$} \\ 
$\frac{N^{\rm bg}}{N^{\rm data}}$ (\%) & {\scriptsize $4.20\pm 0.46$} & {\scriptsize $8.86\pm 0.05$} & {\scriptsize $3.55\pm 0.07$} & {\scriptsize $5.60\pm 0.10$} & {\scriptsize $2.43\pm 0.10$} & {\scriptsize $7.89\pm 0.24$} & {\scriptsize $11.6\pm 1.60$} \\
$N^{\rm sig}$,~$10^3$ & {\scriptsize $3947\pm 7$} & {\scriptsize $1793\pm 5$} & {\scriptsize $319.3\pm 1.8$} & 
{\scriptsize $833.6\pm 7.0$} & {\scriptsize $322.6\pm 4.5$} & {\scriptsize $244.7\pm 3.3$} & {\scriptsize $21.05\pm 1.40$} \\ 
$\varepsilon_{\rm det}$ (\%) & 9.66 & 7.09 & 6.69 & 2.65 & 2.19 & 2.47 & 0.82 \\
$\bigl(\frac{\Delta{\cal B}}{{\cal B}}\bigr)_{\rm syst}$ (\%) & 2.4 & 2.5 & 4.0 & 3.9 & 5.2 & 4.4 & 8.1 \\
\hline 
\end{tabular}
\normalsize
\caption{Results of the event selection, detection efficiency and 
         systematic uncertainties of the branching fractions.}\label{tab:ksdata} 
\end{center} 
\end{table} 
Number of selected events ($N^{\rm data}$) as well as 
background admixture ($N^{\rm bg}/N^{\rm data}$) for 
each decay mode are shown in Table~\ref{tab:ksdata}. 
The main non-$\tau\tau$ background comes from $e^+e^-\to q\bar{q}~(q=u,~d,~s,~c)$ events. 
To take into account cross-feed background six decay modes are analysed simultaneously 
and number of signal events of $i$-th mode is calculated according: 
\[ N^{\rm sig}_{i} = \sum_{j} (\mathcal{E}^{-1})_{ij}(N^{\rm data}_{j} - N^{\rm bg}_{j}),~i=1\div 6,\] 
where $\mathcal{E}_{ij}$ is $6\times 6$ efficiency matrix (diagonal elements of $\mathcal{E}_{ij}$ are shown 
in Table~\ref{tab:ksdata}), $N^{\rm sig}$ numbers for all modes are also summarized in Table~\ref{tab:ksdata}. 

For the $\tau^-\to\pi^-K^0_S\nu_\tau$, $\tau^-\to K^-K^0_S\nu_\tau$, $\tau^-\to\pi^-K^0_S\pi^0\nu_\tau$ and 
$\tau^-\to K^-K^0_S\pi^0\nu_\tau$ modes with large statistics lepton tag is applied and normalisation 
to the two-lepton events $(\tau^{\mp}\to e^{\mp}\nu\nu,~\tau^{\pm}\to\mu^{\pm}\nu\nu)$ (or shortly $e-\mu$) 
is used to calculate branching fractions: 
\[ {\cal B}_{i} = \frac{N^{\rm sig}_{i}}{N^{\rm sig}_{e-\mu}}\frac{{\cal B}_{e}{\cal B}_{\mu}}{{\cal B}_{e}+{\cal B}_{\mu}}, \]
where $N^{\rm sig}_{e \textrm{-}\mu}$ is the number of $e-\mu$ events, ${\cal B}_\ell$ is the branching fraction of 
$\tau^-\to\ell^-\bar{\nu}_\ell\nu_\tau$, $\ell=e,~\mu$. 
To increase statistics for the remaining $\tau^-\to\pi^-K^0_SK^0_S\nu_\tau$ and 
$\tau^-\to\pi^-K^0_SK^0_S\pi^0\nu_\tau$ modes one-prong decay tag and luminosity normalisation method are used: 
\[ {\cal B}_{i} = \frac{N^{\rm sig}_{i}}{2{\cal L}\sigma_{\tau\tau}{\cal B}_{1-\rm prong}},\] 
where $\sigma_{\tau\tau}=(0.919\pm 0.003)$~nb \cite{Banerjee:2007is} is the cross section of $\tau^+\tau^-$ 
production, ${\cal B}_{1-\rm prong}=(85.35 \pm 0.07)\%$ \cite{Beringer:1900zz} 
is the one-prong decay branching fraction. The branching fractions for one inclusive and six 
exclusive $\tau$ hadronic decay modes with $K^0_S$ measured at Belle are summarized in Table~\ref{tab:brres}. 
The branching fractions for $\tau^-\to K^-K^0_S \nu_\tau$ and $\tau^-\to K^-K^0_S\pi^0\nu_\tau$ 
are measured for the first time at the $B$ factories. The results on branching fractions of 
all studied decay modes are consistent with the previous experiments and have better precision. 

Upper limits on the branching fractions of two $\tau$ decay modes with two $K^0_S$ obtained 
recently at $BABAR$ \cite{Lees:2012de} are shown in Table~\ref{tab:brres} as well. 
\begin{table}[htbp]
\begin{center}
\begin{tabular}{@{~}lcl@{~}}
\hline 
\hline 
Mode & Branching fraction & Reference \\ 
\hline
$K^0_S X^- \nu_\tau$           & $(9.15 \pm 0.01 \pm 0.15) \times 10^{-3}$ & Belle \\ 
$\pi^- K^0_S \nu_\tau$         & $(4.16 \pm 0.01 \pm 0.08) \times 10^{-3}$ & Belle \\ 
$K^- K^0_S \nu_\tau$           & $(7.40 \pm 0.07 \pm 0.27) \times 10^{-4}$ & Belle \\ 
$\pi^- K^0_S \pi^0 \nu_\tau$   & $(1.93 \pm 0.02 \pm 0.07) \times 10^{-3}$ & Belle \\ 
$K^- K^0_S \pi^0 \nu_\tau$     & $(7.48 \pm 0.10 \pm 0.37) \times 10^{-4}$ & Belle \\ 
$\pi^- K^0_S K^0_S \nu_\tau$   & $(2.33 \pm 0.03 \pm 0.09) \times 10^{-4}$ & Belle \\ 
$\pi^-K^0_SK^0_S\pi^0\nu_\tau$ & $(2.00 \pm 0.22 \pm 0.20) \times 10^{-5}$ & Belle \\ 
$K^-K^0_SK^0_S\nu_\tau$        & $< 6.3\times 10^{-7}~{\rm at~90\%~CL}$    & $BABAR$ \\    
$K^-K^0_SK^0_S\pi^0\nu_{\tau}$ & $< 4.0\times 10^{-7}~{\rm at~90\%~CL}$    & $BABAR$ \\ 
\hline 
\end{tabular}
\caption{Summary of the branching fractions of the $\tau$ decay modes with 
$K^0_S$ from Belle \cite{Ryu:2014vpc} and $BABAR$ \cite{Lees:2012de}. The first 
uncertainty is statistical and the second one is systematic.}\label{tab:brres} 
\end{center} 
\end{table} 

\subsection{Study of mass spectra in $\tau^-\to\pi^-K^0_SK^0_S\pi^0\nu_\tau$ sample} 

\begin{figure}[hbtp]
\centering 
\includegraphics[width=0.49\textwidth]{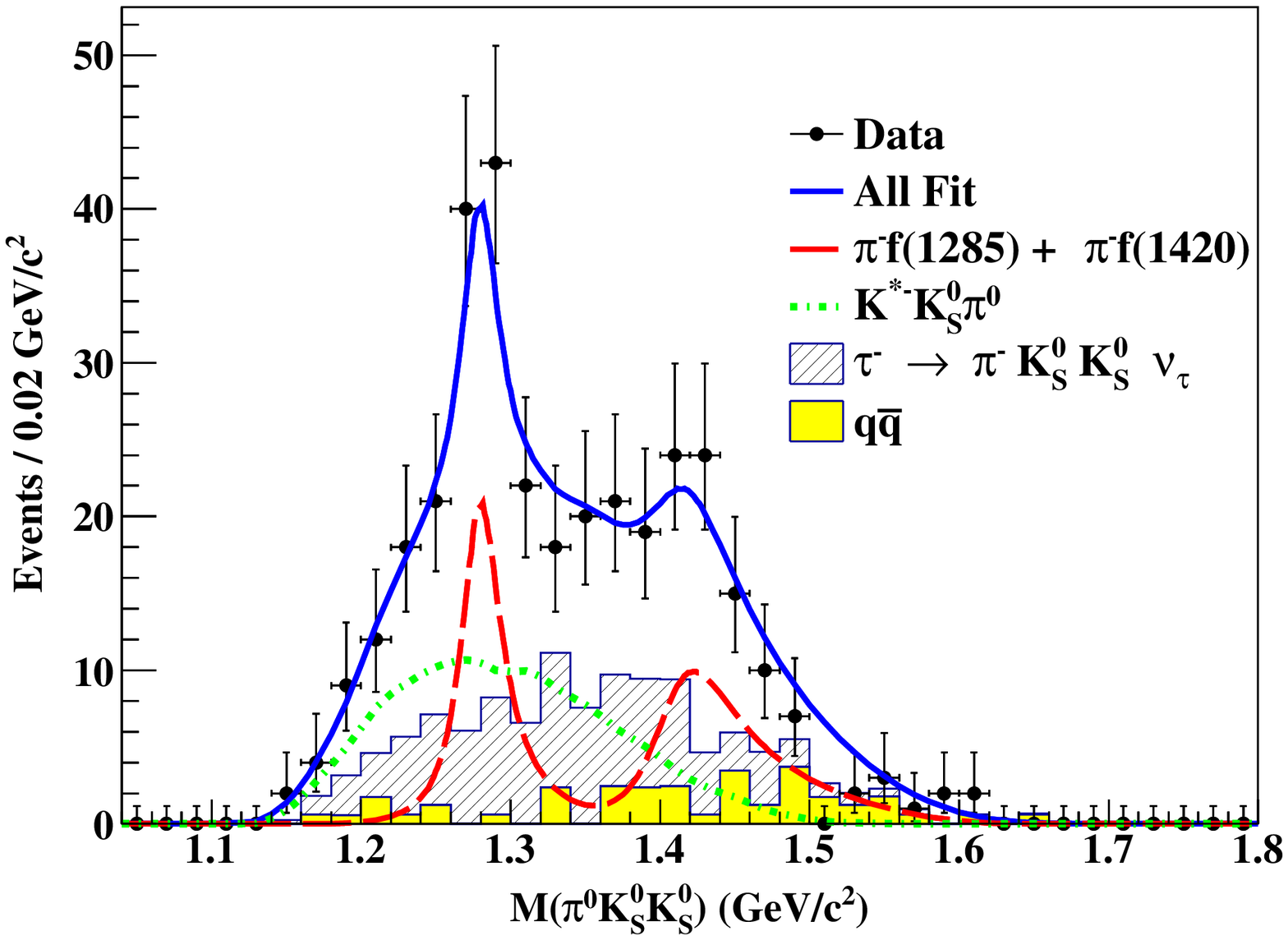} 
\includegraphics[width=0.49\textwidth]{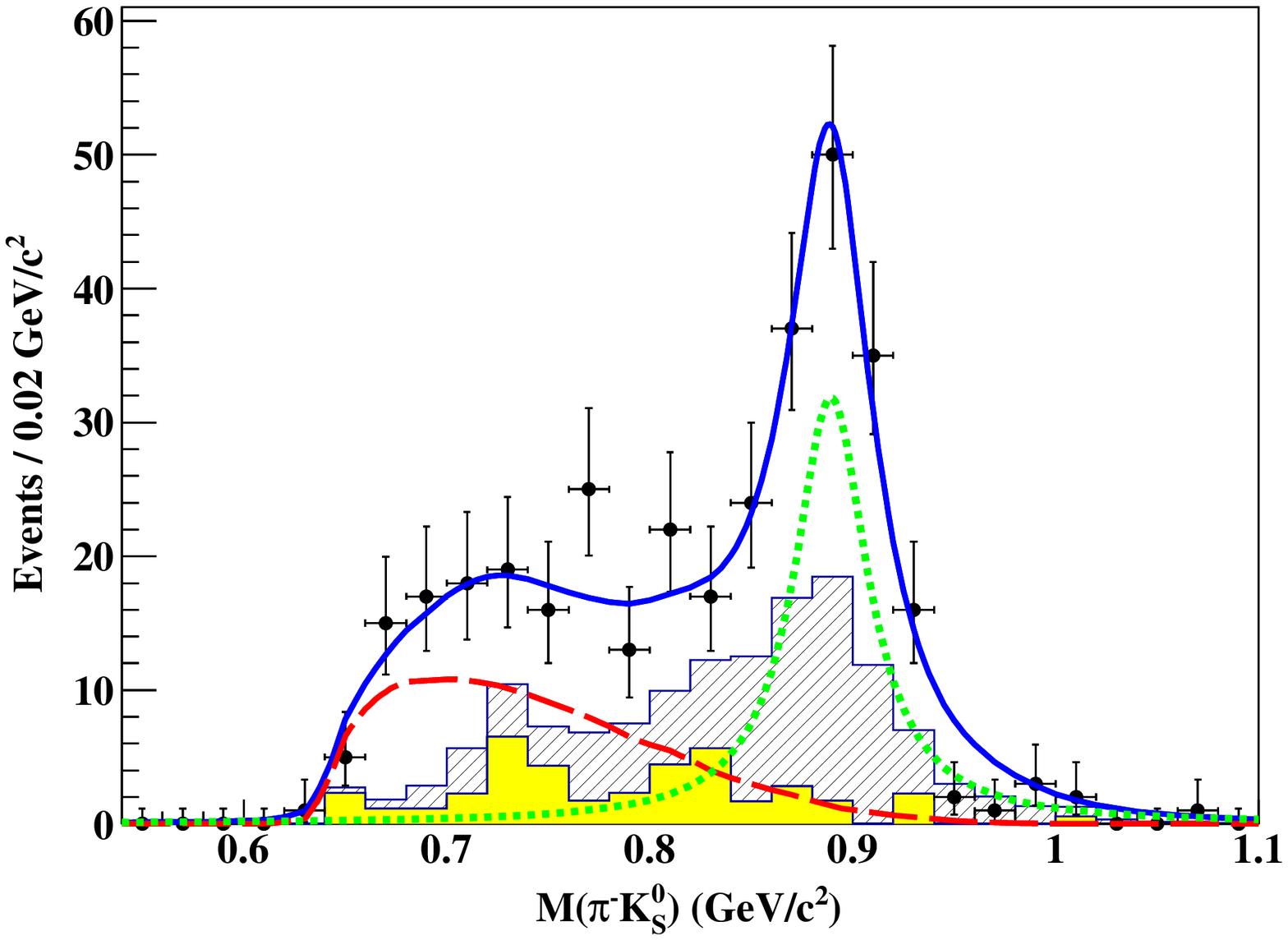}
\\
\caption{Invariant mass of the $K^0_S K^0_S \pi^0$ (left) and $\pi^- K^0_S$ (right) 
subsystems for the $\tau^-\to\pi^-K^0_SK^0_S\pi^0\nu_\tau$ events. Points with errors 
are experimental data, the hatched histogram is background from 
$\tau^-\to\pi^- K^0_S K^0_S \nu_\tau$, and the shaded (yellow) histogram is the 
continuum $q\bar{q}$ background. Solid line is the result of the fit by the 
$\tau^-\to (f_1(1285)\pi^-~+~f_1(1420)\pi^-~+~K^{*-}K^0_S)\,\nu_\tau$ 
model for signal events. The $(f_1(1285)\pi^-~+~f_1(1420)\pi^-)\,\nu_\tau$ 
and $K^{*-} K^0_S \nu_\tau$ contributions are shown by the dashed (red) 
and dotted (green) lines respectively.}\label{ksdynam} 
\end{figure}

The invariant masses of the $K^0_S K^0_S \pi^0$ and $\pi^- K^0_S$ subsystems 
for $\tau^-\to\pi^-K^0_SK^0_S\pi^0\nu_\tau$ events are shown in Fig.~\ref{ksdynam}. 
Clear peak at the mass of about $1280$~MeV/$c^2$ is seen in the $K^0_S K^0_S \pi^0$ 
mass distribution as well as some bump around $1420$~MeV/$c^2$, taking into account the 
quantum numbers of the hadronic final state observed structures are associated with 
the intermediate $f_1(1285)$($J^{PC}=1^{++}$) and $f_1(1420)$($J^{PC}=1^{++}$) 
pseudovector mesons. The $\pi^- K^0_S$ distribution exhibits clear signal from the 
intermediate $K^{*-}(892)$ vector meson. No additional resonancelike structures 
are observed in the other mass spectra. 
To determine the observed contributions simultaneous fit of the $M(K^0_S K^0_S \pi^0)$ 
and $M(\pi^- K^0_S)$ mass spectra is performed. The result of the fit is also shown 
in Fig.~\ref{ksdynam}. 
The fractions (significances) of three contributions in $\tau^-\to\pi^-K^0_SK^0_S\pi^0\nu_\tau$ 
are extracted to be $(34\pm 5)\%$ ($12\sigma$), $(12\pm 3)\%$ ($4.8\sigma$) and $(54\pm 6)\%$ ($7.8\sigma$) 
for the $f_1(1285)\pi^-\nu_\tau$, $f_1(1420)\pi^-\nu_\tau$ and $K^{*-}(892)K^0_S\pi^0\nu_\tau$ mechanisms, 
respectively. With obtained fractions, products of the branching fractions for the subprocesses 
are calculated (the first uncertainty is statistical and the second one is systematic): 
\begin{eqnarray*}
{\cal B}(\tau^-\to f_1(1285)\pi^-\nu_\tau)\cdot{\cal B}(f_1(1285)\to K^0_S K^0_S\pi^0)  & = & (0.68\pm 0.13\pm 0.07)\times 10^{-5}, \\ 
{\cal B}(\tau^-\to f_1(1420)\pi^-\nu_\tau)\cdot{\cal B}(f_1(1420)\to K^0_S K^0_S\pi^0)  & = & (0.24\pm 0.05\pm 0.06)\times 10^{-5}, \\ 
{\cal B}(\tau^-\to K^*(892)^-K^0_S\pi^0\nu_\tau)\cdot{\cal B}(K^*(892)^-\to K^0_S\pi^-) & = & (1.08\pm 0.14\pm 0.15)\times 10^{-5}. 
\end{eqnarray*}

\section{Search for lepton flavor violating $\tau$ decays at LHCb} 

In the analysis the statistics of about 1~fb$^{-1}$ collected by LHCb 
at the proton-proton c.m.s. energy $\sqrt{s}=7$~TeV in 2011 is used. 
The inclusive $\tau^-$ production cross section at LHCb is 
$\sigma_{\rm inc}(\tau^-)=80~\mu$b (80\% of $\tau^-$ come from $D^-_s\to\tau^-\bar{\nu}_\tau$ decay).  
LFV decay $\tau^-\to\mu^-\mu^+\mu^-$ and the LNV and BNV ($|\Delta({\rm B-L})|=0$) 
decays $\tau^-\to \bar{p}\mu^+\mu^-$ and $\tau^-\to p\mu^-\mu^-$ were studied. 
The decay $D^-_s\to\phi\pi^-$ followed by $\phi\to\mu^+\mu^-$ 
(or shortly $D^-_s\to\phi(\mu^+\mu^-)\pi^-$), which has similar 
experimental signature is used for the calibration and normalization. 

The following selection criteria are applied: 
good tracks with transversal momenta satisfying $p^{\rm 1trk}_{T}>0.3$~GeV/$c$ ; 
the transversal momentum of the combined three-track system 
is required to be $p^{\rm 3trk}_{T}>4$~GeV/$c$; decay length of 
the $\tau$ candidate satisfy $\lambda_\tau >$100~$\mu$m; angle 
between 3-track momentum and radius vector of $\tau$ decay 
vertex should be small; invariant mass of $\mu^+\mu^-$ 
satisfy $M_{\mu^+\mu^-}>0.45$~GeV/$c^2$ to suppress background 
from the $D^-_s\to\eta(\to\mu^+\mu^-\gamma)\mu^-\bar{\nu}_\mu$ decay. 

\begin{figure}[htbp]
\begin{center}
\begin{minipage}[b]{0.32\linewidth}
	\centering
	\begin{overpic}[width=\textwidth]{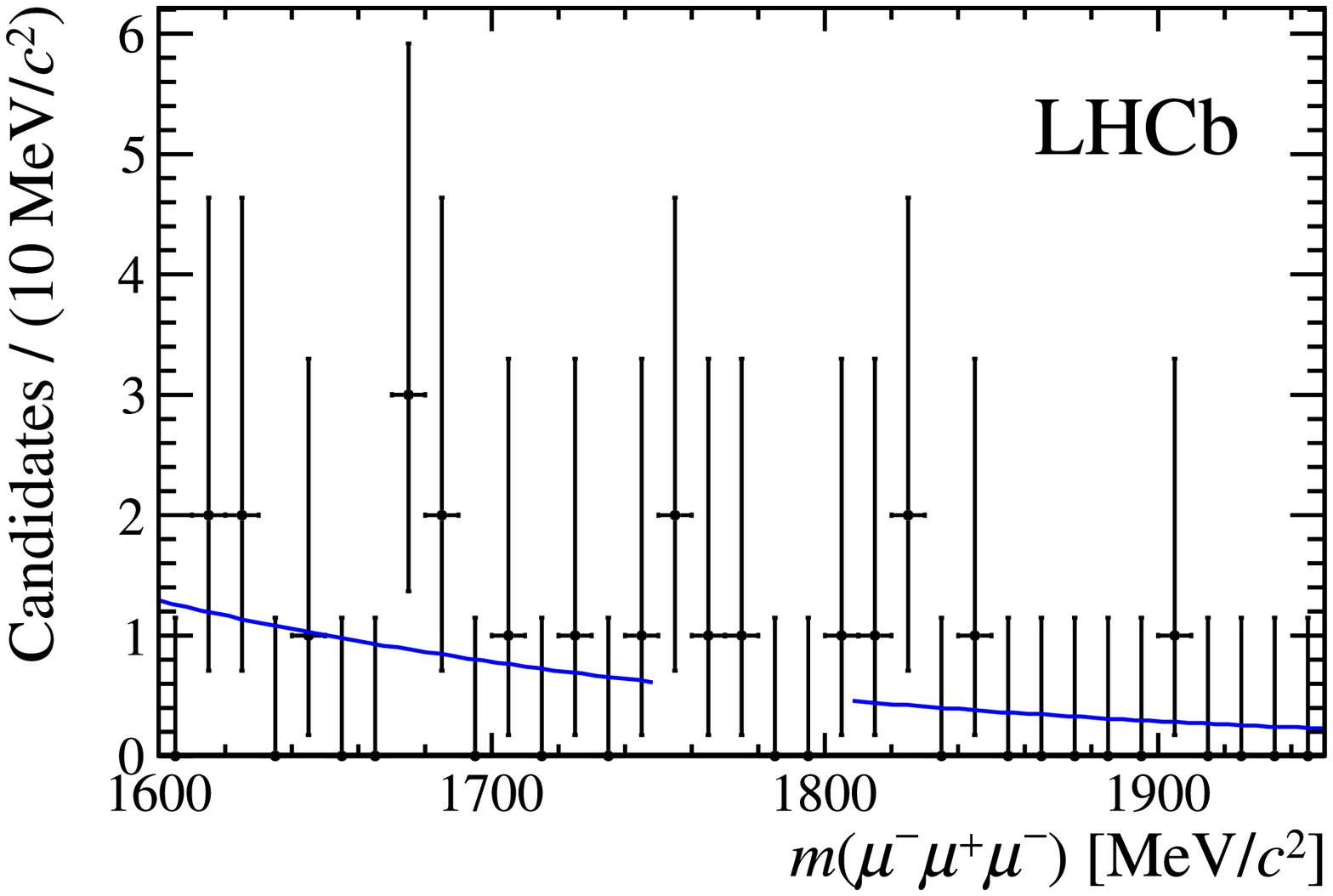}
	\put (20,75) {\scriptsize{\bf(a)}}
	\put (35,80) {\tiny{${\cal M}_{\rm 3body} \in  [0.65, 1.0]$}}
	\put (35,70) {\tiny{${\cal M}_{\rm PID}   \in [0.725, 1.0]$}}
	\end{overpic}
\end{minipage}
\begin{minipage}[b]{0.32\linewidth}
	\centering
	\begin{overpic}[width=\textwidth]{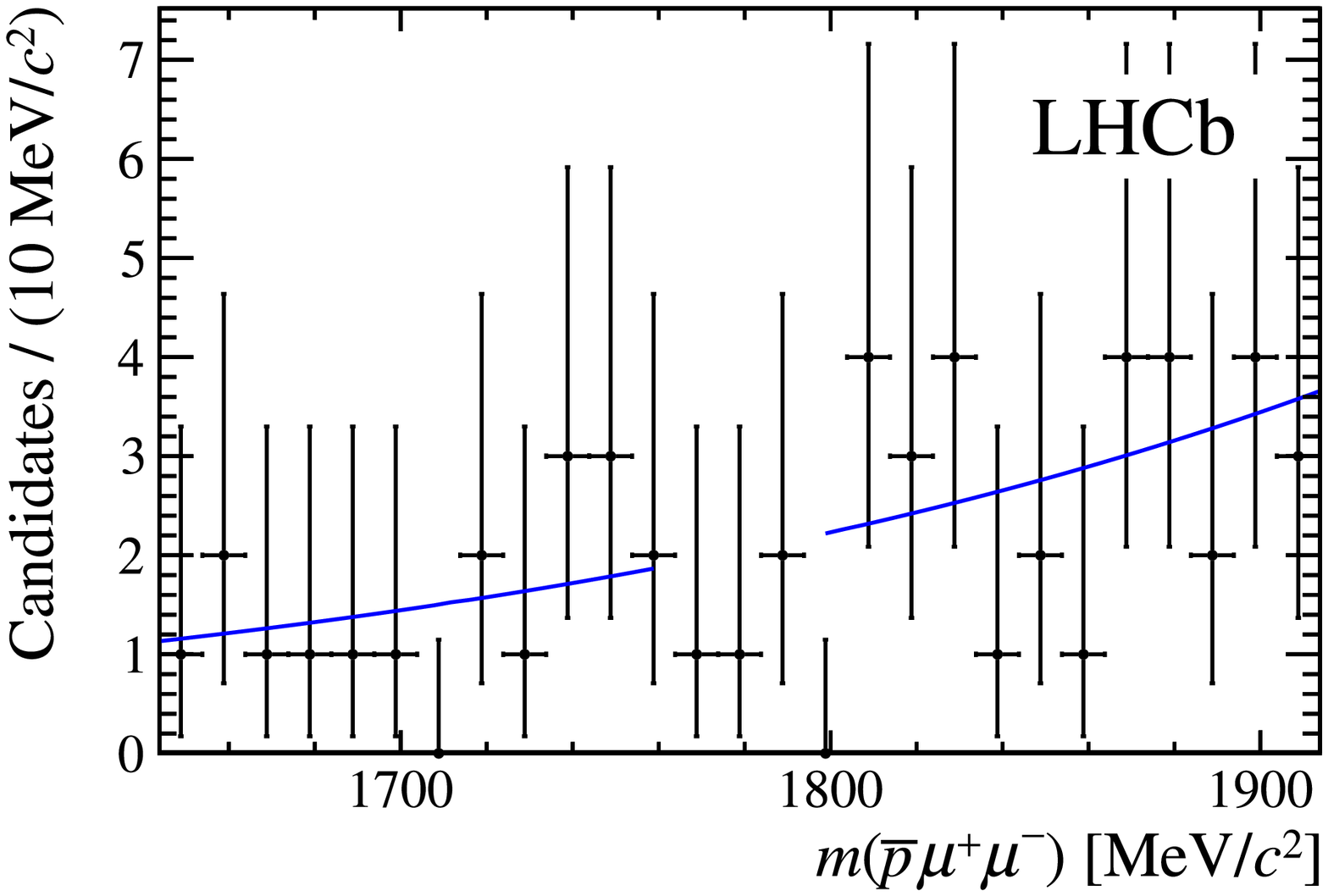}
	\put (20,75) {\scriptsize{\bf(b)}}
	\put (35,77) {\tiny{${\cal M}_{\rm 3body} \in [0.40, 1.0]$}}
	\end{overpic}
\end{minipage}
\begin{minipage}[b]{0.32\linewidth}
	\centering
	\begin{overpic}[width=\textwidth]{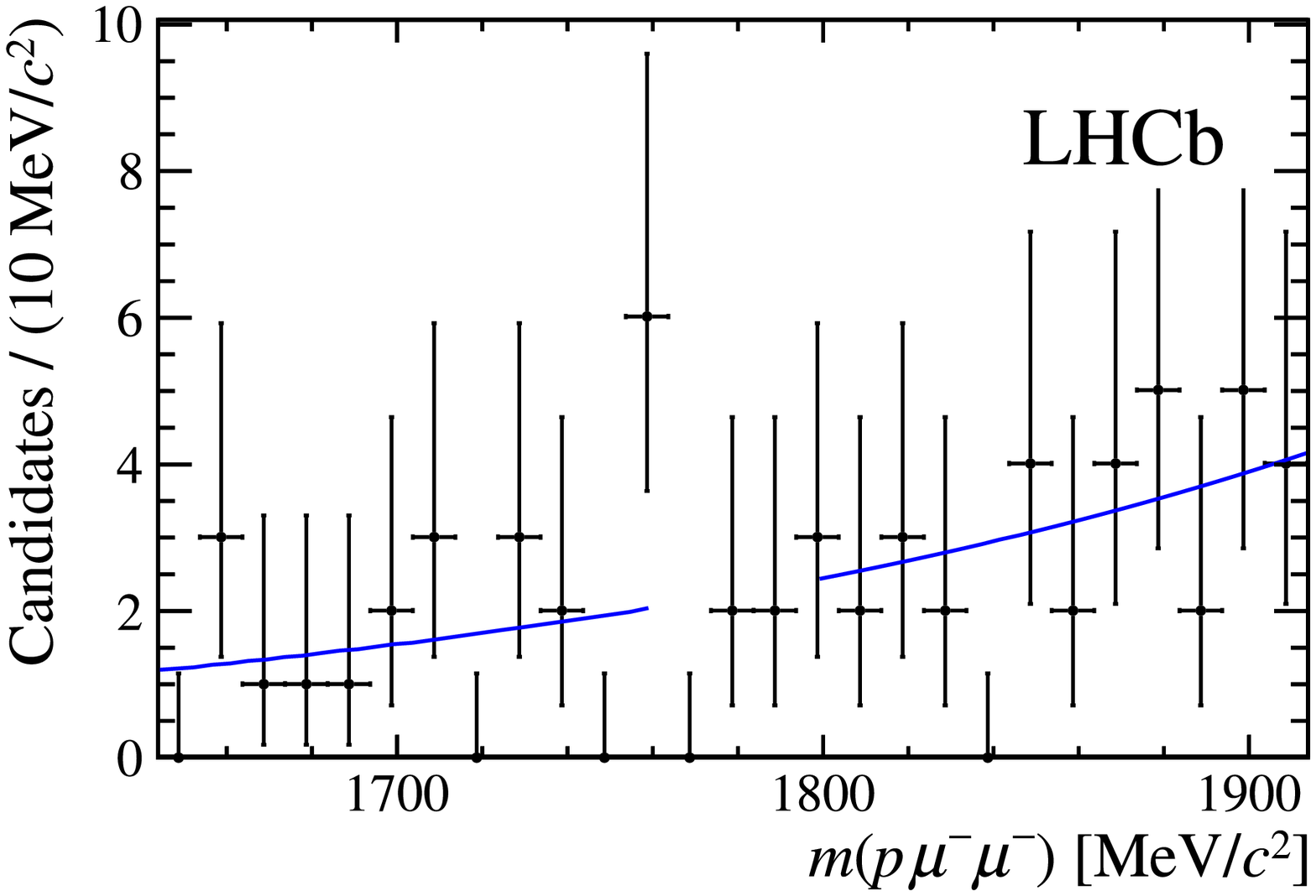}
	\put (20,75) {\scriptsize{\bf(c)}}
	\put (33,77) {\tiny{${\cal M}_{\rm 3body} \in [0.40, 1.0]$}}
	\end{overpic}
\end{minipage}
\caption{Invariant mass distributions and fits 
of the mass sidebands in data for: 
(a) $\tau^-\to\mu^-\mu^+\mu^-$ candidates, 
(b) $\tau^-\to\bar{p}\mu^+\mu^-$ candidates, and 
(c) $\tau^-\to p\mu^-\mu^-$ candidates.}\label{lfv:fit}
\end{center}
\end{figure}

\begin{figure}[hbtp]
\centering 
\includegraphics[width=\textwidth]{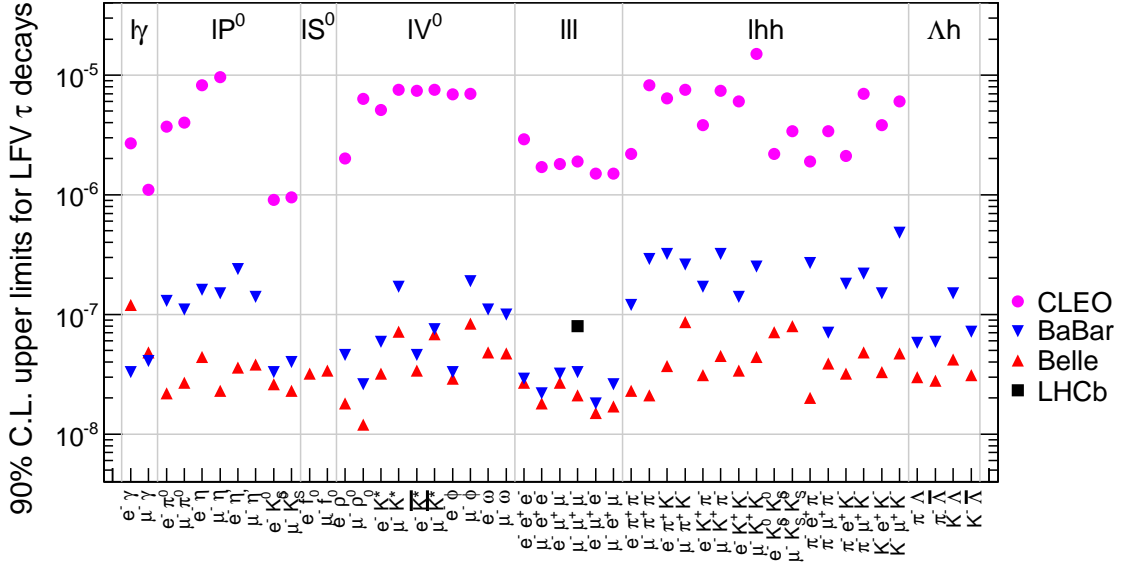} 
\caption{Summary of the upper limits (90\%CL) on the branching fractions of 48 LFV $\tau$ decays 
measured at $B$ factories.~Recent result on ${\cal B}(\tau^-\to\mu^-\mu^+\mu^-)$ from LHCb is shown 
by black square.}\label{lfvres} 
\end{figure}

For the further analysis the region around $\tau$ mass 
$(m_{\tau}-20$~MeV/$c^2) <{\cal M}_{\rm inv}<(m_{\tau}+20$~MeV/$c^2)$ ($\pm 2\sigma_M$) 
is blinded (so called blinded analysis). 
After all selections each $\tau$ candidate is characterized 
by a probability to be signal or background according to the 
values of three likelihood parameters: 
${\cal M}_{\rm 3body}$ includes geometrical properties to 
identify displaced 3-body $\tau$ decays, ${\cal M}_{\rm PID}$ is particle 
identification based on the information from the detector subsystems: RICH, ECAL, Muon \cite{Alves:2008zz}, 
${\cal M}_{\rm inv}$ is invariant mass of $\tau$ decay products. 
The range for each classifier, ${\cal M}_{\rm 3body}$ (varies from -1 to 1) and ${\cal M}_{\rm PID}$ (varies from 0 to 1), 
is subdivided into 6 and 5 unequally spaced bins respectively. 
And ${\cal M}_{\rm inv}$ within $\pm 20$~MeV/$c^2$  mass window around $\tau$ mass is subdivided 
into 8 equally spaced bins to extract number of signal events. 
Result of the fit of the data sidebands for particular bins in ${\cal M}_{\rm 3body}$ and ${\cal M}_{\rm PID}$ 
for the three decay modes are shown in Fig.~\ref{lfv:fit}.
The branching fraction of signal decay is calculated according: 
\[ {\cal B}_{\rm sig}={\cal B}(D^-_s\to\phi(\mu^+\mu^-)\pi^-)\frac{f^{D_s}_\tau}{{\cal B}(D^-_s\to\tau^-\bar{\nu}_\tau)}\frac{\epsilon^{\rm det}_{\rm norm}\epsilon^{\rm trg}_{\rm norm}}{\epsilon^{\rm det}_{\rm sig}\epsilon^{\rm trg}_{\rm sig}}\frac{N_{\rm sig}}{N_{\rm norm}},\] 
where $N_{\rm sig}$($N_{\rm norm}$) is number of selected signal (normalisation) events, 
$\epsilon^{\rm trg}_{\rm sig},~\epsilon^{\rm det}_{\rm sig}$ ($\epsilon^{\rm trg}_{\rm norm},~\epsilon^{\rm det}_{\rm norm}$) 
are trigger and detection efficiency for signal (normalisation) events, $f^{D_s}_\tau=0.78\pm 0.05$ is fraction of 
$\tau$'s from $D^-_s$ decays. 
The obtained upper limits on the branching fractions: 
\begin{eqnarray*}
{\cal B}(\tau^-\to\mu^-\mu^+\mu^-)    & < & 8.0\times 10^{-8}~{\rm at}~90\%~{\rm CL}, \\
{\cal B}(\tau^-\to\bar{p}\mu^+\mu^-)  & < & 3.3\times 10^{-7}~{\rm at}~90\%~{\rm CL}, \\
{\cal B}(\tau^-\to p\mu^-\mu^-)       & < & 4.4\times 10^{-7}~{\rm at}~90\%~{\rm CL}.
\end{eqnarray*}
Figure~\ref{lfvres} summarizes the upper limits on the branching fractions of 48 
LFV modes obtained at $B$ factories. After the LHCb upgrade \cite{Lucchesi:2014gra} it will be possible 
to improve the upper limit on ${\cal B}(\tau^-\to\mu^-\mu^+\mu^-)$ in comparison 
with the current best limit obtained by Belle \cite{Hayasaka:2010np}. 
For the first time the upper limits on the branching fractions of 
the LNV and BNV $\tau^-\to \bar{p}\mu^+\mu^-$ and $\tau^-\to p\mu^-\mu^-$ 
decays were measured at LHCb. 

\section{CP violation in $\tau$ decays at $B$ factories}
Recent studies of CPV in the $\tau^-\to\pi^- K_S(\geq\pi^0)\nu_{\tau}$ decays 
at $BABAR$~\cite{BABAR:2011aa} as well as in the $\tau^-\to K_S\pi^-\nu_{\tau}$ decay 
at Belle~\cite{Bischofberger:2011pw} provide complementary information about sources 
of CPV in these hadronic decays. 

The decay-rate asymmetry $A_{\rm CP}=\frac{\Gamma(\tau^+\to\pi^+ K_S(\geq\pi^0)\nu_{\tau})-\Gamma(\tau^-\to\pi^- K_S(\geq\pi^0)\nu_{\tau})}{\Gamma(\tau^+\to\pi^+ K_S(\geq\pi^0)\nu_{\tau})+\Gamma(\tau^-\to\pi^- K_S(\geq\pi^0)\nu_{\tau})}$ 
was studied at $BABAR$ with the $\tau^+\tau^-$ data sample of $\int Ldt=476$~fb$^{-1}$. 
The obtained result $A_{\rm CP}=(-0.36\pm 0.23\pm 0.11)\%$ is about $2.8$ standard deviations 
from the SM expectation $A^{K^0}_{\rm CP}=(+0.36\pm 0.01)\%$.

At Belle CPV search was performed as a blinded analysis based on a $699~{\rm fb}^{-1}$ data sample. 
Specially constructed asymmetry, which is a difference between the mean values of $\cos\beta\cos\psi$ for 
$\tau^-$ and $\tau^+$ events, was measured in bins of $K^0_S\pi^-$ mass squared ($Q^2=M^2(K^0_S\pi)$): 
\[ A_i^{CP}(Q^2_i) = \frac{\int\limits_{\Delta Q^2_i}\cos{\beta} \cos{\psi}  \left( \frac{d\Gamma_{\tau^-}}{d\omega} - \frac{d\Gamma_{\tau^+}}{d\omega} \right) d\omega}{\frac{1}{2} \int\limits_{\Delta Q^2_i}\left(\frac{d\Gamma_{\tau^-}}{d\omega}+\frac{d\Gamma_{\tau^+}}{d\omega} \right)d\omega} \simeq \langle \cos\beta\cos\psi \rangle_{\tau^-} - \langle \cos\beta\cos\psi \rangle_{\tau^+}, \] 
where $\beta$, $\theta$ and $\psi$ are the angles, evaluated from the measured parameters of the final hadrons, 
$d\omega=dQ^2 d\!\cos\theta d\!\cos\beta$. In contrary to the decay-rate asymmetry the introduced 
$A_i^{CP}(Q^2_i)$ is already sensitive to the CPV effects from the charged scalar boson exchange \cite{Kuhn:1996dv}. 
No CP violation was observed and the upper limit on the CPV 
parameter $\eta_S$ was extracted $|{\rm Im}(\eta_S)|<0.026$~at~90\% CL. Using this limit parameters 
of the Multi-Higgs-Doublet models \cite{Grossman:1994jb,Choi:1994ch} can be constrained as 
$|{\rm Im}(XZ^*)|<0.15~M_{H^\pm}^2/(1\,\mathrm{GeV}^{2}/c^4)$, where 
$M_{H^\pm}$ is the mass of the lightest charged Higgs boson, the complex constants Z and X 
describe the coupling of the Higgs boson to leptons and quarks respectively. 
 
The other more complicated and most powerful method to extract 
CPV parameter at $e^+ e^-$ $B$ factories is an unbinned maximum likelihood 
fit of events in the full phase space. The main idea of this method is to consider 
events where both taus decay to the particular final states. 
One $\tau^{\mp}$ (signal side) decays to particular hadronic final state 
(for example $\tau^\mp\to (K\pi)^\mp \nu$) and the other $\tau^{\pm}$ 
(tag side) decays to some well investigated mode with large branching fraction. 
As a tag decay mode we can take $\tau^{\pm}\to\pi^{\pm}\pi^0\nu$ 
having the largest branching fraction, it also serves as spin analyser, 
which allows one to be sensitive to the spin dependent part of the differential 
decay width of signal decay using effect of spin-spin correlation of taus \cite{Tsai:1971vv}. 
In the technique we analyze $e^+ e^-\to \tau^{\mp}\tau^{\pm}\to ((K\pi)^\mp\nu,~\pi^{\pm}\pi^0\nu)$ 
(or shortly $((K\pi)^\mp,~\pi^{\pm}\pi^0)$) events in the 12-dimensional 
phase space. The probability density function is constructed from the total differential 
cross section, which is a sum of spin independent term and spin-spin correlation term. 

To write the total differential cross section we follow the approach developed in \cite{Fetscher:1990su,Tamai:2003he}. 
The differential cross section of $e^+e^-\to\tau^+(\vec{\zeta}^{*+})\tau^-(\vec{\zeta}^{*-})$ 
reaction in the center-of-mass system (c.m.s.) is given by formula \cite{Tsai:1971vv}: 
\[ \frac{d\sigma(\vec{\zeta}^{*-},\vec{\zeta}^{*+})}{d\Omega} = 
\frac{\alpha^2}{64E^2_{\tau}}\beta_{\tau}(D_0+D_{ij}\zeta^{*-}_i \zeta^{*+}_j) \] 
\[ D_0 = 1+\cos^2{\theta}+\frac{1}{\gamma^2_{\tau}}\sin^2{\theta} \]  
\begin{equation}
D_{ij} = \left( \begin{array}{@{}c@{~~}c@{~~}c@{}}
(1+\frac{1}{\gamma_{\tau}^2})\sin^2{\theta} & 0 & \frac{1}{\gamma_{\tau}}\sin{2\theta} \\
 0 & -\beta^2_{\tau}\sin^2{\theta} & 0 \\ 
\frac{1}{\gamma_{\tau}}\sin{2\theta} & 0 & 1+\cos^2{\theta}-\frac{1}{\gamma^2_{\tau}}\sin^2{\theta} \\ 
\end{array} \right)
\label{eqtotdif}
\end{equation}
where $\vec{\zeta}^{*\mp}$ is polarisation vector of $\tau^{\mp}$
in the $\tau^{\mp}$ rest frame (unitary vector along $\tau^{\mp}$ spin direction). 
Asterisk marks parameters measured in the associated $\tau$ rest
frame. $\alpha$, $E_{\tau}$, $\gamma_{\tau}=E_{\tau}/M_{\tau}$, 
$\beta_{\tau}=P_{\tau}/E_{\tau}$ and $\theta$ are fine structure constant, energy, Lorentz
factor, velocity of $\tau$ (in the units of $c$) and polar angle of 
$\tau^-$ momentum direction, respectively. 
Signal differential decay width is written in the form: 
\[\frac{d\Gamma(\tau^\mp(\vec{\zeta}^*)\to (K\pi)^\mp\nu)}{dm^2_{K\pi}d\Omega^*_{K\pi}d\tilde{\Omega}_{\pi}}=\begin{array}{c} (A_0+{\eta_{CP}}A_1)+(\vec{B}_0+{\eta_{CP}}\vec{B}_1)\vec{\zeta}^* \\ (A_0+{\eta^*_{CP}}A_1)-(\vec{B}_0+{\eta^*_{CP}}\vec{B}_1)\vec{\zeta}^* \end{array},\] 
where $\eta_{CP}$ is CPV sensitive parameter; $m_{K\pi}$, $\Omega^*_{K\pi}$ are $K\pi$ 
invariant mass and solid angle of the $K\pi$ system in the $\tau$ rest frame, 
$\tilde{\Omega}_\pi$ is solid angle of the charged pion in the $K\pi$ rest frame. 
$A_1$ and $A_2$ are form factors of the spin-independent part; $\vec{B}_0$ and $\vec{B}_1$ are 
form factors of the spin-dependent part of the differential decay width.  

The $\tau^{\pm}(\vec{\zeta'}^*)\to\rho^{\pm}(K^*)\nu(q^*)\to\pi^{\pm}(p^*_1)\pi^0(p^*_2)\nu(q^*)$ 
decay width reads (with unimportant for this analysis total normalization constant $\kappa_{\rho}$): 
\begin{equation}
\frac{d\Gamma(\tau^{\pm}\to\pi^{\pm}\pi^0\nu)}{dm^2_{\pi\pi}d\Omega^*_{\rho}d\tilde{\Omega}_{\pi}}=A'\mp\vec{B}'\vec{\zeta'}^*, 
\end{equation}
\[ Q^*=p^*_1-p^*_2,~K^*=p^*_1+p^*_2,~m^2_{\pi\pi}=K^{*2}, \] 
\[ A'=\kappa_\rho W(m^2_{\pi\pi})(2(q,Q)Q^*_0-Q^2q^*_0),~\vec{B}'=\kappa_\rho W(m^2_{\pi\pi})(Q^2\vec{K}^*+2(q,Q)\vec{Q}^*),\] 
\[ W(m^2_{\pi\pi})=|F_{\pi}(m^2_{\pi\pi})|^2\frac{p^*_{\rho}(m^2_{\pi\pi})\tilde{p}_{\pi}(m^2_{\pi\pi})}{M_{\tau}m_{\pi\pi}},~p^*_{\rho}=\frac{M_{\tau}}{2}\biggl(1-\frac{m^2_{\pi\pi}}{M^2_{\tau}}\biggr),\] 
\begin{equation} 
\tilde{p}_{\pi}=\frac{\sqrt{(m^2_{\pi\pi}-(m_{\pi}+m_{\pi^0})^2)(m^2_{\pi\pi}-(m_{\pi}-m_{\pi^0})^2)}}{2m_{\pi\pi}}, 
\end{equation} 
where $p^*_{\rho}$, $\Omega^*_{\rho}$ are momentum and solid angle of $\rho$ meson in the $\tau$ rest frame, 
$\tilde{p}_{\pi}$, $\tilde{\Omega}_{\pi}$ are momentum and solid angle of charged pion in the $\rho$ rest frame, 
$F_{\pi}(m^2_{\pi\pi})$ is pion form factor \cite{Fujikawa:2008ma}. 

As a result the total differential cross section for $((K\pi)^\mp,\rho^{\pm})$ events can be written as \cite{Tsai:1971vv}: 
\begin{equation}
\frac{d\sigma((K\pi)^{\mp},\rho^{\pm})}{dm^2_{K\pi}d\Omega^*_{K\pi} d\tilde{\Omega}_\pi dm^2_{\pi\pi}d\Omega^*_\rho d\tilde{\Omega}_\pi d\Omega_\tau}=\frac{\alpha^2\beta_{\tau}}{64E^2_{\tau}}\biggl({\cal F}+\begin{array}{c} {\eta_{CP}} \\ {\eta^*_{CP}} \end{array}{\cal G}\biggr), 
\end{equation} 
\[ {\cal F}=D_0A_0A'-D_{ij}B_{0i}B'_j,~{\cal G}=D_0A_1A'-D_{ij}B_{1i}B'_j.\] 
Experimentally we measure particle parameters in the c.m.s.,  
hence visible differential cross section is given by: 
\scriptsize
\begin{equation}
\frac{d\sigma((K\pi)^{\mp},\rho^{\pm})}{dp_{K\pi}d\Omega_{K\pi}dm^2_{K\pi}d\tilde{\Omega}_{\pi}dp_{\rho}d\Omega_{\rho}dm^2_{\pi\pi}d\tilde{\Omega}_{\pi}}=\sum\limits_{\Phi_1,\Phi_2}\frac{d\sigma((K\pi)^{\mp},\rho^{\pm})}{dm^2_{K\pi}d\Omega^*_{K\pi} d\tilde{\Omega}_\pi dm^2_{\pi\pi}d\Omega^*_\rho d\tilde{\Omega}_\pi d\Omega_\tau}\biggl|\frac{\partial (\Omega^*_{K\pi},\Omega^*_{\rho},\Omega_{\tau})}{\partial (p_{K\pi},\Omega_{K\pi},p_{\rho},\Omega_{\rho})}\biggr|, 
\end{equation}
\normalsize
where the summation is done over the unknown $\tau$ direction, which 
is determined with two-fold ambiguity by $\Phi_1$ and $\Phi_2$ angles. 
They are calculated using parameters measured in the experiment \cite{Tamai:2003he}. 

As a result $\eta_{\rm CP}$ can be extracted in the simultaneous unbinned maximum 
likelihood fit of the $((K\pi)^-,\rho^+)$ and $((K\pi)^+,\rho^-)$ events in the 
12-dimensional phase space. The advantage of this method is that we use the 
whole information recorded in the experiment to extract CPV parameter. More 
over this, with this method it is possible to study CPV in the spin-dependent 
part of the decay width, i.e.~we can test NP models with the nontrivial $\tau$ 
spin-dependent effects. Similar method is used at $e^+ e^-$ $B$ factories to 
measure Michel parameters in leptonic $\tau$ decays \cite{epifanov}.

\end{document}